\documentclass[aps,prl,showpacs,twocolumn,floats]{revtex4}
\usepackage{amssymb}

\usepackage{bm}
\usepackage[latin1]{inputenc}
\usepackage[T1]{fontenc}
\usepackage{amsmath,amssymb}
\usepackage{amsthm}
\usepackage{amscd,pst-all}
\usepackage{graphicx,color}
\usepackage{hyperref}

\newcommand{\mb}[1]{{\ensuremath{\mathbf{#1}}}}
    {\endverbatim\endquote\vspace{0pt}}

\begin{document}

\title{Ultra-fast spin dynamics: the effect of colored noise.}
\author{U.Atxitia$^1$, O.\ Chubykalo-Fesenko$^1$, R.W.Chantrell$^2$, U.Nowak$^3$, and A Rebei$^4$ }
\date{\today}

\affiliation{$^1$ Instituto de Ciencia de Materiales de Madrid,
CSIC, Cantoblanco, 28049 Madrid, Spain} \affiliation{$^2$ Department
of Physics, University of York, York YO10 5DD, U.~K.}
\affiliation{$^3$ Fachbereich Physik, Universit\"at Konstanz, 78457 Konstanz, Germany}
\affiliation{$^4$ Seagate Research, 1251 Waterfront place,
Pittsburgh, PA 15222, U.~S.~A.}

\begin{abstract}
Recent experimental results have pushed the limits of magnetization dynamics to pico- and femtosecond timescales. This ultra-fast spin dynamics occurs in extreme conditions of strong and rapidly varying fields and high temperatures.  This situation requires new description of magnetization dynamics, even on a phenomenological level of the atomistic Landau-Lifshitz-Gilbert equation, taking into account that the correlation time for electron system could be of the order of the inverse characteristic spin frequency. For this case we introduce the thermodynamically correct phenomenological approach for spin dynamics based on the Landau-Lifshitz-Miyasaki-Seki equation.  The influence of the noise correlation time on  longitudinal and transverse magnetization relaxation is  investigated. We also demonstrate the effect of the noise correlation time on  demagnetisation rate of different materials during laser-induced dynamics.
\end{abstract}

\maketitle



One of the fundamental questions of modern solid state physics is
how rapidly the magnetization can respond to an external excitation.
The recent development of  time-resolved pump-probe experimental techniques using X-ray
spectroscopy based on synchrotron radiation \cite{Siegmann} and the Stanford
linear accelerator (SLAC) \cite{Tudosa, Stamm}  has allowed the
investigation of magnetization dynamics  on the pico-second time scale. The use of the  powerful femto-second
lasers \cite{beaurepairePRL96, koopmansPRL00, Gerrits} has pushed this limit down to the
femto-second time-scale. The physical processes
 underlying the response of the magnetization on this ultra-short time-scale are complicated and
 far from being understood, but clearly involve the excitation and consequent
 non-equilibrium interaction of electron, phonon and spin sub-systems. Spin dynamic processes on
  this time-scale occur under
 extreme conditions remarkably different from those typical for dynamics at longer time-scales.
 Firstly, the three subsystems (electron, phonon and spin) are not in equilibrium with each other.
 Secondly, spin dynamic processes occur under very strong fields with different sources. In particular,
 in the experiments using the SLAC the magnitude of the external field can be as large as 20T.
The strongly non-homogeneous magnetization processes are driven by the exchange field, having
 a
 magnitude
 $\gtrsim 100$T.
 Finally,  in the
  laser-induced magnetization dynamics the  effective temperature is increased up to and often
  above the Curie temperature \cite{koopmansPRL00,beaurepairePRL96}.

Atomistic spin models have proved to be a powerful approach to model
ultra-fast magnetization dynamics \cite{kazantsevaEPL08, kazantsevaPRB08}. For example, in the
case of laser-induced magnetization changes, spin models provide
important physical insight into the spin-reordering process,
establishing the linear character of the demagnetization during the
sub-picosecond regime and predicting the origin of different
recovery rates in the pico-second regime. The basis of these models
is the stochastic Landau-Lifshitz (LL) equation for each
localized magnetic moment $\mb{s}_i$:
 \begin{equation}
 \label{LLG}
    \mb{\dot{s}}_i=\gamma[\mb{s}_i\times\mb{H}_i]-\gamma\alpha[\mb{s}_i\times[\mb{s}_i\times\mb{H}_i]]
\end{equation}
Here $\mb{H}_i$ is the local effective field which includes Zeeman,
exchange, anisotropy and magnetostatic contributions, augmented with
a stochastic field $\mb{\xi}_i(t)$ with the following properties for
both components and different spin sites:
\begin{equation}
\label{whitenoise}
     \left\langle \xi_i(t) \right\rangle=0, \ \  \    \left\langle\xi_i(t)\xi_j(t')\right\rangle=
     \frac{2\alpha k_B T}{\gamma \mu_{s}}\delta(t-t')\delta_{ij}.
\end{equation}
 Here $T$ is the temperature, $\gamma$ is the gyro-magnetic ratio, $\mu_{s}$ is the magnetic moment  and
 $\alpha$ is the  parameter describing the coupling to the bath system.
 The basis of this equation is the separation of timescales, assuming that the bath
  (phonon or electron system) is much faster than the spin system. In this case, the bath degrees of freedom can be averaged out and replaced by a stochastic field with white noise correlation functions. The coefficient in front of the delta function in Eq.(\ref{whitenoise})  is determined by the fluctuation-dissipation theorem. The assumption of white noise is therefore invalid for magnetization dynamics occurring on a timescale comparable to the
 the relaxation time of the electron system.  The typical correlation time for the electron system in metals is $\lesssim 10 fs$ \cite{Siegmann}. Such magnetization dynamics time scale is now commonly achieved by applying femtosecond laser pulses. A further limitation of this approach comes
    from the fact that  characteristic frequencies of the magnetization process are now also of the
    order  of the timescale of the noise (electron) variable. Therefore, for modelling the ultra-fast
    magnetization experiments the approach (\ref{LLG}) could break down.

 The aim of the present Letter is to present a  classical formalism beyond the white noise approximation
  to form a strong physical basis for models of ultra-fast
 magnetization dynamics in extreme conditions. First, we introduce
 the formalism  and show that our approach is consistent with the equilibrium
 Boltzmann distribution and coincides with the previous atomistic approach for small correlation
 times for the bath variable.
 As main implications of the correlated noise approach, we discuss the influence of noise correlations on the most relevant characteristics of
  magnetization dynamics: the longitudinal
 and transverse relaxation times. Finally, we model the laser-induced demagnetisation rate for materials with different noise correlation times.

 The standard generalization of the white noise to include correlations is  the
 Ornstein-Uhlenbeck  stochastic process \cite{Sancho}. However, when we implemented this process within the LL dynamics, we have
seen that in agreement with the general theory \cite{Hasegawa},  the Boltzmann distribution at equilibrium is not
recovered for magnetization dynamics with correlated noise in this case. The deviations invariably correspond to precessional
frequencies of the order of the inverse correlation time. The colored
noise approach based on the Ornstein-Uhlenbeck process  may provide
a reasonable description in some situations when the system goes to
a stationary condition not necessary coinciding with the equilibrium
one (an example of this could be the spin-torque pumping into a
magnetic system). However, in experiments such as those
corresponding to laser-pulsed induced dynamics, a stochastic
approach giving the correct equilibrium magnetization long after the
laser pulse is gone, is necessary.

A suitable approach has been found in the work of Miyazaki and Seki
\cite{MSeki} who generalized the Langevin equation for one spin to
a non-Markovian case. The approach has been introduced for one spin
at high temperatures, neglecting the interactions with other spins
and assuming that their role is to provide the bath environment. In
the present paper we generalize this approach to a many spin case,
similar to the standard way of Eq.(\ref{LLG}) where the applied
field is substituted by the local field. We assume that the bath
variable is
due to external sources such as electrons. The other assumption made
in this approach is that the spin is connected locally to the bath.
Consequently, the set of equations for magnetization dynamics (in
the following called Landau-Lifshitz-Miyazaki-Seki (LLMS)) reads:

\begin{eqnarray}
\label{MS}
\mathbf{\dot{s}}_i&=&\gamma [\mathbf{s}_i \times (\mathbf{H}_i + \boldsymbol{\eta}_i)], \nonumber \\
\boldsymbol{\dot{\eta}}_i&=&-{1 \over \tau_c}(\boldsymbol{\eta}_i - \chi \mathbf{s}_i)+ \mathbf{R}_i
\end{eqnarray}
with the fluctuation-dissipation theorem for the bath variable:
$\left\langle\mathbf{R}_i(t)\right\rangle=0$ ; $\left\langle\mathbf{R}_i(t)\mathbf{R}_j(t')\right\rangle=(2\chi k_B T /
\tau_c)\delta_{ij}\delta(t-t')$. The parameter $\chi$ describes
the coupling of the bath variable to the spin. The precession term in
the first equation of the set (\ref{MS})  has the same form as in
the Eq.(\ref{LLG}). However, the damping is now described by the
second equation in this set where also the bath variable
adjusts to the direction of the spin due to the interaction with it.
In the limit $\tau_c
\rightarrow 0$  the stochastic LL equation (\ref{LLG}) is
recovered \cite{MSeki}. This also provides a relation between the damping  and
the coupling constants as $\alpha=\gamma \chi \tau_c $, giving
a more precise physical sense to the LL damping constant at
atomistic level.

\begin{figure}[h]
  \begin{center}
   \includegraphics[scale=1]{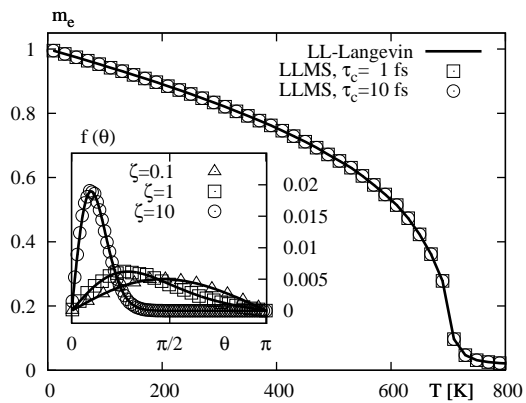}
  \end{center}
  \caption{Equilibrium magnetization as a function of temperature for a system of $\mathcal{N}=32^3$ interacting spins, integrating the LLMS equation with different
   correlation times   and integrating the LL equation.
  The inset shows distribution functions for non-interacting spin system modeled within the LLMS approach for different values of the reduced field
  $\zeta=\mu_s H/k_B T$ ($\gamma H \tau_c=1.76$) and correlation time $\tau_c=10$fs. The solid line in the inset represents the Boltzmann distribution.
  \label{f:LLMS}}
\end{figure}

For integration of Eqs.(\ref{MS}) the Heun integration scheme
 was modified for this special case.
   First of all we investigated the equilibrium properties  for an ensemble of non-interacting spins.  In all cases of large fields, temperatures
   and correlation times, the correct Boltzmann distribution is
   obtained at equilibrium (see inset in Fig.\ref{f:LLMS}).

Now we turn to the multi-spin system.   First of all, we prove that
the stochastic Eqs.(\ref{MS}) for a multi-spin system are consistent
with the standard equilibrium properties. For this purpose we use
the formalism of the Onsager kinetic coefficient method applied in
Ref.\cite{Brownian} for the LL multispin system (\ref{LLG}). The system
(\ref{MS}) is linearized near equilibrium and represented in a
general form of the Langevin equation:
\begin{eqnarray}
\label{genLan}
{dx_i \over dt}&=&-\sum_j \gamma_{ij}X_j + r_i\\ \nonumber
\left\langle r_i(t)\right\rangle &=&0;\;\;\left\langle r_i(0)r_j(t)\right\rangle =\mu_{ij}\delta(t)
\end{eqnarray}
Here $x_i$ stands for  small deviations of the stochastic variables
$\mathbf{s}_i$ or $\eta_i$ from their equilibrium values, $X_i$ represent their
thermodynamically conjugate variables and
$\mu_{ij}=\gamma_{ij}+\gamma_{ji}$. For the spin variable we have:
$X_j=-(\mu_s / k_B T) H_j$ , where $H_j$ is the internal
field corresponding to a particular lattice site and spin component.
Unlike Eq.(\ref{LLG}), the first equation in Eqs.(\ref{MS}) contains
only a precessional term and, therefore, the corresponding kinetic
coefficients are antisymmetric in spin components, giving for this
equation $\mu_{ij}=0$. Taking into account the generalization of the
internal energy to include the bath variable as
$F(\{\mathbf{s}_i\},\{\mathbf{\eta}_i\})
=F_0(\{\mathbf{s}_i\})+\sum_{i}[\eta_i^2/(2
\chi)-\mathbf{\eta}_i\mathbf{s}_i]$, where
$F_0(\{\mathbf{s}_i\})$ is the internal energy without the bath
variable, the conjugate variable to the bath one is
$\mathbf{X}_j=(\mathbf{\eta}_j-\chi \mathbf{s}_j)/(k_B T \chi)$.
Therefore, the corresponding matrix of the kinetic coefficients is
diagonal and for the second equation we obtain $\mu_{ij}= (2k_BT
\chi /T_c) \delta_{ij}$. Consequently, we have proven that, under
the supposition of local coupling of  the spin to the bath
variables, the set of multi-spin equations (\ref{MS}) is consistent
with the equilibrium properties.

 In our simulations for the multispin system we use a Heisenberg Hamiltonian on a cubic system of $32^3$ magnetic moments with nearest-neighbor interactions only,
with $\mu_s=1.45 \mu_B$ and the Curie temperature $T_c=700$K ($k_B
T_c \approx 1.44$J). The coupling
parameter $\chi$ was chosen to give the LL damping parameter
$\alpha=0.01$.  In Fig.\ref{f:LLMS} we present
calculations of the equilibrium magnetization as a function of
temperature for spin systems with different values of the
correlation times. Independence of the equilibrium properties on the
correlation time, and the agreement with calculations using the LL
equation with uncorrelated noise demonstrates our generalization of
the LLMS equation to multi-spin systems. Consequently, the LLMS
equation provides a basis for the phenomenological description of
magnetization dynamics in extreme situations of high temperatures,
 and large and rapidly varying external fields. The advantage of the approach is
also that the fluctuation-dissipation theorem is not applied
directly to the spin variable. Therefore, the bath variable (for
example electrons) and the spin system need not be in equilibrium with each
other.

 Next we discuss the most important implications of this new approach to the ultra-fast dynamics.
 It is known that during the excitation with spatially inhomogeneous fields in the Terahertz
 range \cite{Siegmann,Stamm} and also during laser-induced  magnetization dynamics \cite{beaurepairePRL96,koopmansPRL00},
 strong local disordering of the spin system occurs.   The dynamics in these cases is governed
 by field or temperature excited high-frequency spin waves which are responsible for the
 effective damping. The important dynamical feature then is the rate of magnetization recovery.
 During these processes two types of relaxation could be distinguished. The first one
 known as longitudinal relaxation is responsible for linear magnetization recovery, i.e. the
 magnetization magnitude.
 During the laser-induced demagnetization, the longitudinal relaxation is responsible for
 the femtosecond demagnetization.
 The second one is the transverse relaxation when the magnetization vector relaxes to the
 direction parallel to the internal field via magnetization precession. The longitudinal
 relaxation time increases with the temperature while the transverse relaxation time has
 minimum at $T_c$ \cite{chubykaloPRB06}.

\begin{figure}[t]
  \begin{center}
   \includegraphics[scale=0.7]{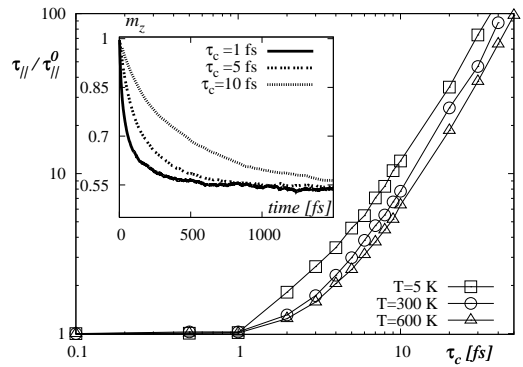}
  \end{center}
  \caption{Longitudinal relaxation time (normalized to the uncorrelated case) as a function of the correlation time for various temperatures calculated within LLMS approach. The inset shows longitudinal relaxation for various correlation times and $T=600$K.
  \label{f:Long}}
\end{figure}

 To simulate the longitudinal relaxation, we start with
the initial condition $\{s_i^z\}=1.0$ and observe the system to
relax at given temperature $T$. The obtained relaxation curves are
then fitted to exponential decay to extract the longitudinal
relaxation rate. The longitudinal relaxation time, normalized to the
uncorrelated case, is presented in Fig.\ref{f:Long} as a function of
the noise correlation time. The longitudinal time calculated by
means of the LL approach (\ref{LLG}) is of the order of $10$fs
($\tau^{0}_{||}= 28$fs  at $T=300$K). For correlation time $\tau_c \lesssim 1$fs the uncorrelated
approach gives the same results as the LLMS one.  However, one can
see that $\tau_c \simeq 10-100$fs gives a dramatic increment of the
longitudinal relaxation time. The effect is less pronounced at
higher temperature since in this case the temperature contributes to
the loss of correlations.

\begin{figure}[h]
  \begin{center}
   \includegraphics[scale=0.7]{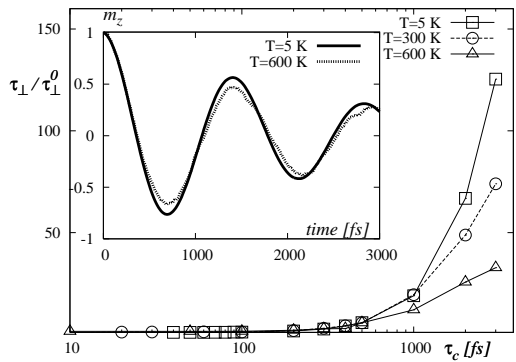}
  \end{center}
  \caption{Transverse relaxation time (normalized to the uncorrelated case) as a function of the correlation time for various temperatures calculated within LLMS approach. The insert shows transverse relaxation for two temperatures $T=5$K and $T=600$K and $\tau_c=10$fs.
  \label{f:Trans}}
\end{figure}

Next, we investigate the transverse relaxation in Fig.\ \ref{f:Trans}. The transverse relaxation time is defined by the magnetisation precession and normally is  much slower than the
longitudinal one. For one spin  $\tau_{\perp}=\tau_{\perp}^0 [1+(\omega_H \tau_c)^2]$ \cite{MSeki}, where $\omega_H$ is the field-dependent precessional frequency.  Consequently, the influence of the correlation time on the transverse relaxation may be expected only for strong applied field
 for which $\omega_{H}\sim \tau_c^{-1}$ and, thus,  could be relevant for SLAC experiments. To show the influence of the noise correlation on the transverse relaxation, we model the precessional dynamics at strong applied field $H=24.85$T ($0.05J/\mu_s$). In this case, the spin system was first equilibrated at given temperature and applied field.  After that the whole system was rotated to an angle $30^o$ and the relaxation to the direction parallel to the applied field was observed.
 For this particular strong applied field,  the correlation times $\tau_c \gtrsim 100fs$  are necessary in order to see their influence on the transverse relaxation.

\begin{figure}[h]
  \begin{center}
   \includegraphics[scale=1.5]{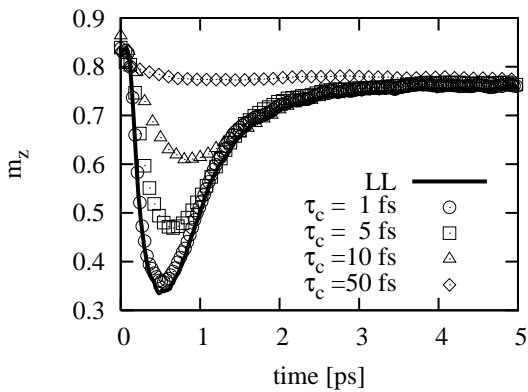}
  \end{center}
  \caption{Laser-induced demagnetisation as a function of time for various noise correlation times $\tau_c$ modelled within LLMS and LL approaches.
  \label{f:pump_probe}}
\end{figure}

 To demonstrate the immediate physical consequence of the correlated noise for a more concrete experimental situation, we present in Fig.\ref{f:pump_probe} the modelling results for the laser-induced demagnetisation for various noise correlation times and constant damping parameter $\alpha=0.05$. We suppose that the
magnetization dynamics is produced by a Gaussian laser pulse with 50 fs duration and the fluency $33.7 mJ/cm^2$. Similar to Ref.\cite{kazantsevaEPL08}, we assume that the photon energy is transfered to the electrons and lattice but
the magnetization is directly coupled to the electron temperature $T_e$. The latter is calculated within the two-temperature model \cite{...}
with the electron and lattice specific heat constants $C_e= 700 J/m^3 K^2 \times T_e^{max},\; C_l=3\times 10^6 J/m^3K $ and the coupling constant
$G_{el}=8\times10^{17} W/m^3 K $.  Our results clearly show strong impact of the noise correlation time on the degree of demagnetisation during the laser-induced process. Namely, the materials with small $\tau_c$ are demagnetised easier. This could be true, for example, for  $d$-electrons in metals with large scattering rate or for $f$-electrons in rare-earths which have strongly  relativistic nature.

In conclusion, the standard phenomenological approach to model spin
dynamics has been generalized to the non-Markovian case. This
approach is necessary in the extreme situations of large
characteristic magnon frequencies occurring during ultra-fast
magnetization processes. The advantages of the new approach are the
following:  (i) the memory (correlation) effects arising from the
fact that the bath variable responds to the spin direction are taken
into account. This corresponds to the situation when the bath
variable is not in equilibrium with the spin system. (ii) the
fluctuation-dissipation theorem is not applied to the spin systems
as in the standard LL approach  (iii) In equilibrium the
Boltzmann distribution is recovered. The price for this new approach
is the use of two phenomenological constants: the phenomenological
damping parameter $\alpha$ for the LL approach is substituted by
two phenomenological parameters in the LLMS approach: the
correlation time $\tau_c$ and the coupling constant $\chi$.
Several processes may be important in determining these constants, as for example, the spin-orbit coupling,
momentum relaxation, scattering rate and de-phasing time of conduction electrons.
As in the LL approach, these parameters will be material-specific and
their physical origins should be clarified on the basis of
first-principle approaches. We have shown that the ultra-fast magnetization dynamics is strongly influenced by these parameters
which stresses the necessity of first-principle models, capable to
clarify  their physical origins.

The authors acknowledge the financial support from Seagate Technology, USA and from European COST-P19 Action.



\begin{thebibliography}{99}

\bibitem{Siegmann}  J. St\"ohr and  H. C. Siegmann,  "\textit{Magnetism}" (Springer-Verlag, Berlin, 2006) and references therein.



\bibitem{Tudosa} I.Tudosa, C.Stamm, A.B.Kashubam F.King, H.C.Siegmann, J.J. St\"ohr, G.Ju, B.Lu and D.Weller, Nature {\bf 428}, 831 (2004)

\bibitem{Stamm} C. Stamm, I. Tudosa, H. C. Siegmann, J. St\"ohr, A. Yu. Dobin, G. Woltersdorf, B. Heinrich and A. Vaterlaus, Phys. Rev. Lett.
\textbf{94} 197603 (2005).

\bibitem{beaurepairePRL96} E. Beaurepaire, J.-C. Merle, A. Daunois,
and J.~Y. Bigot, Phys. Rev. Lett. \textbf{76}, 4250 (1996).

\bibitem{koopmansPRL00}
B. Koopmans, M. van Kampen, J.~T. Kohlhepp, and W.~J.~M. de~Jonge, Phys. Rev.
  Lett. {\bf 85},  844  (2000).

\bibitem{Gerrits}
Th. Gerrits, H.A.M. van der Berg, J.Hohlfeld, L. B\"ur and Th.Rasing, Nature {\bf 418} 509 (2002)




\bibitem{kazantsevaEPL08}
N. Kazantseva, U. Nowak, R.~W. Chantrell, J. Hohlfeld, and A. Rebei, Europhys.
  Lett. {\bf 81},  27004  (2008).

\bibitem{kazantsevaPRB08} N.Kazantseva, D.Hinzke, U.Nowak, R.W.Chantrell, U.Atxitia and O.Chubykalo-Fesenko, Phys Rev B {\bf 77}, 184428 (2008)


\bibitem{Sancho} J. M. Sancho  and M. San Miguel,  Phys. Rev. A \textbf{26}, 1589 (1982).



\bibitem{Hasegawa} H. Hasegawa, Phys.A \textbf{387} 2697 (2008).




\bibitem{MSeki} K. Miyazaki and K. Seki, J. Chem. Phys. \textbf{108}, 7052 (1998)

\bibitem{Brownian} O. A. Chubykalo, R. Smirnov-Rueda, J. M. González,
M. A. Wongsam, R. W. Chantrell and U. Nowak, J. Magn. Magn. Mater. \textbf {266},
28 (2003).

\bibitem{chubykaloPRB06}
O. Chubykalo-Fesenko, U. Nowak, R.~W. Chantrell, and D. Garanin, Phys. Rev. B
  {\bf 74},  094436  (2006).








\end{thebibliography}
\end{document}